\documentclass[pra,preprint,showpacs]{revtex4}
\usepackage[centertags]{amsmath}
\usepackage[koi8-r]{inputenc}
\usepackage{amsfonts}
\usepackage{amssymb}
\usepackage{amsthm} 
\usepackage{newlfont}
\usepackage{epsfig}
\usepackage{amscd}
\usepackage{graphicx}
\usepackage{epsfig}
\usepackage{footnote}
\usepackage{lipsum}
\usepackage{showlabels}
\usepackage{color}
\usepackage{xcolor}
\usepackage{graphicx}
\usepackage{multirow}
\usepackage{subcaption}
\usepackage{amscd}

\newcommand{\beq}{\begin{equation}}
\newcommand{\eeq}{\end{equation}}
\newcommand{\ba}{\begin{array}}
\newcommand{\ea}{\end{array}}
\newcommand{\bea}{\begin{eqnarray}}
\newcommand{\eea}{\end{eqnarray}}
\begin{document}

\begin{center}
{\large \sc \bf Remote restoring of $(0,1)$-excitation states and concurrence scaling}

\vskip 15pt

N.A.Tashkeev$^1$ and A.I.Zenchuk$^2$


\vskip 8pt

{\it $^1$Lomonosov Moscow State Uniersity, Moscow, 119991, Russia}

{\it $^2$Federal Research Center of Problems of Chemical Physics and Medicinal Chemistry RAS,
Chernogolovka, Moscow reg., 142432, Russia}.

\end{center}

\begin{abstract}
We study the long distance (0,1)-excitation state restoring in the linear open chain  governed by the XX-Hamiltonian. We show that  restoring  the 1-order coherence matrix results  in restoring the 1-excitation block of the 0-order coherence matrix, so that only one  0-excitation element of the density matrix remain unrestored. Such restoring  also scales the concurrence between any two  qubits of the transferred state, the scaling factor is defined by the Hamiltonian and doesn't depend on the initial sender's state. 
Sender-Receiver entanglement is also studied via the PPT criterion.
\end{abstract}


\maketitle

\section{Introduction}
\label{Section:Introduction}
Quantum state transfer is one of the most  attractive processes to be studied in  quantum informatics. Since the  celebrated paper by Bose \cite{Bose}, where this problem was first formulated, many achievements have been reached in improving the quality of the state transfer, first of all, fidelity of state transfer. Two method resoving this problem were invented  withing several years after the above paper. These are the perfect state transfer, which can be achieved in the completely inhomogeneous $XX$-chain with nearest neighbor interaction  \cite{CDEL,KS}, and the high-fidelity state transfer along the homogeneous spin chain  with remoted end-nodes \cite{GKMT}. It was shown later that namely the spin chain with properly remote end-nodes serves as a model providing  stability  of state transfer under small perturbations of the Hamiltonian  
\cite{CRMF,ZASO,ZASO2,ZASO3}. As for practical realization, the photon systems seem to be most effective for long distance communication 
\cite{PBGWK,PBGWK2,DLMRKBPVZBW}. However, a spin chain is a quite reasonable candidate for the short distance state transfer \cite{PSB,LH}. 

Another approach to the problem of state transfer, called the  optimized state transfer, is applicable mainly to the mixed state transfer. In that process, the elements of the receiver's density matrix $r$ are proportional to the appropriate elements of the sender's density matrix $s$ \cite{FPZ_2021} (so-called state restoring):
\begin{eqnarray}\label{restore}
r_{ij} = \lambda_{ij} s_{ij}, \;\;\;\forall \; i,j,
\end{eqnarray}
up to some diagonal elements which can not be given the  form (\ref{restore}) because of the trace-normalization condition. 
The above  structure of $r$  can be obtained due to applying the special unitary transformation to the so-called extended receiver  (several nodes of a spin chains including the spins of the receiver itself). The task is simplified by using the evolution under the Hamiltonian preserving the number of excited spins \cite{FZ_2017}, when the Hamiltonian $H$ describing  evolution of an $N$-qubit quantum system can be represented in the following block-diagonal form:
\begin{eqnarray}\label{Hd}
H={\mbox{diag}}(H^{(0)},H^{(1)},\dots,H^{(N)}),
\end{eqnarray}
where each block $H^{(j)}$ governs the $j$-excitation state subspace. In particular, $H^{(0)}$ and $H^{(N)}$ are scalars because there is only one $0$-excitation state and one $N$-excitation state in the $N$-qubit system. Then the spin dynamic yields independent evolution of  multi-quantum coherence matrices which reduces mixing  different matrix elements during evolution. By the $n$-order multi-quantum coherence matrix $\rho^{(n)}$ we mean the set of such  elements of the density matrix that are responsible  for transitions in the state-space of a quantum system with changing the  $z$-projection of the total spin momentum by $n$. Here $z$ is some selected direction. For instance, in the case of a system in the strong external homogeneous magnetic field, $z$ is the direction of the magnetic field. Thus, the  density matrix $\rho$ of the $N$-qubit system can be written as the following sum:
\begin{eqnarray}\label{mq}
\rho=\sum_{n=-N}^N \rho^{(n)}.
\end{eqnarray}

We have to emphasize, that there is a principal difference in restoring the higher order coherence matrices $\rho^{(n)}$, $|n|>0$, and 0-order coherence matrix $\rho^{(0)}$. The matter is that $\rho^{(0)}$ must satisfy the trace-normalization, ${\mbox{Tr}}\rho^{(0)}=1$, which makes impossible the complete restoring of 0-order coherence matrix, as indicated above. It was shown in \cite{Z_2018} that only non-diagonal elements of $\rho^{(0)}$ can be restored via the unitary transformation of the extended receiver. This forced us to introduce different tool for treating $\rho^{(0)}$ in Ref.\cite{BFLP_2022}, where we   showed the existence of such  0-order coherence matrix of the sender $\rho^{(S;0)}$ that can be perfectly transferred to the receiver. We emphasize that the effect of perfect transfer can not be realized for the higher order coherence matrices. 

In our paper, we restrict the spin dynamics to the evolution of the  0- and 1-excitation state-subspaces which we call $(0,1)$-excitation state space. This  allows to obtain some specific results as features of state space restricted in this way. To organize such evolution, we have to start with  the initial state of the quantum system including only $0$- and $1$-excitation states and use the Hamiltonian in the block-diagonal form (\ref{Hd}). The basis of such subspace for an $N$-qubit quantum system  reads
\begin{eqnarray}\label{basis 1}
|n\rangle,\;\;\;n=0,1,\dots,N,
\end{eqnarray}
where $n$ is the number of a single excited spin and $|0\rangle$ means the state without excitations. 
The density matrix has the following structure
\begin{eqnarray}\label{rho}
\rho= \left(
\begin{array}{c|c}
\rho^{(0)}_{00}& \rho^{(1)}_{0I}\cr\hline
\rho^{(-1)}_{I0}&\rho^{(0)}_{II}\end{array}
\right),
\end{eqnarray}
where $\rho^{(1)}_{0I}$ is the $1\times N$ 1-order coherence matrix responsible for the transitions from $0$-excitation to $1$-excitation subspaces;  $\rho^{(-1)}_{I0}$ is the  $N\times 1$ (-1)-order coherence matrix responsible for transitions from the (-1)-excitation to 0-excitation subspace; $\rho^{(0)}_{00}$ and $\rho^{(0)}_{II}$ compose $0$-order coherence matrix, where $\rho^{(0)}_{00}$ (a scalar) is the probability of $0$-excitation state and $\rho^{(0)}_{II}$ collects the transitions amplitudes between the 1-excitation states. The explicit form of $\rho$ reads
\begin{eqnarray}\label{rhoexpl}
\rho= \left(
\begin{array}{c|ccc}
\rho^{(0)}_{00}& \rho^{(1)}_{01}&\cdots& \rho^{(1)}_{0N} \cr\hline
\rho^{(-1)}_{10}&\rho^{(0)}_{11} &\cdots&\rho^{0}_{1N}\cr
\cdots & \cdots & \cdots & \cdots \cr
\rho^{(-1)}_{N0}&\rho^{(0)}_{N1} &\cdots&\rho^{0}_{NN}\end{array}
\right).
\end{eqnarray}
  Below we demonstrate that restoring the 1-order coherence matrix also restores the block $\rho^{(0)}_{II}$ of the  0-order coherence matrix. We also discover a particular feature of the considered state-restoring  associated with  a special behavior of the concurrence \cite{HW,Wootters} as an acknowledged measure of quantum correlations. Namely, the ratio of 
  the concurrence between any two spins of the Sender and the concurrence between the corresponding spins of the Receiver is defined by the evolution operator and is independent on the initial sender's state.

The structure of the paper is following. In Sec.\ref{Section:SpinDynamics} we describe the (0,1)-excitation spin dynamics, the (0,1)-excitation multi-qubit  state transfer and introduce  the unitary transformation  of the extended receiver. Sec.\ref{Section:StateRestoring} is devoted to restoring the (0,1)-excitation state of the receiver. Entanglement in the  restoring process is studied in Sec.\ref{Section:Entanglement} using both the Wootters (bi-qubit entanglement) and PPT (sender-receiver entanglement) criteria \cite{Peres,VW}. Some examples of state restoring together with entanglement characteristics are given in Sec.\ref{Section:Examples}. Basic results are discussed in Sec.\ref{Section:Conclusions}.

\section{Spin dynamics}
\label{Section:SpinDynamics}

{We consider the communication line consisting of  the sender $S$,   receiver $R$ embedded into the  extended receiver $ER$, and   transmission line  $TL$ connecting the sender to the receiver, see Fig.\ref{Fig:line}.
\begin{figure}[t]
\includegraphics[ scale=0.4]{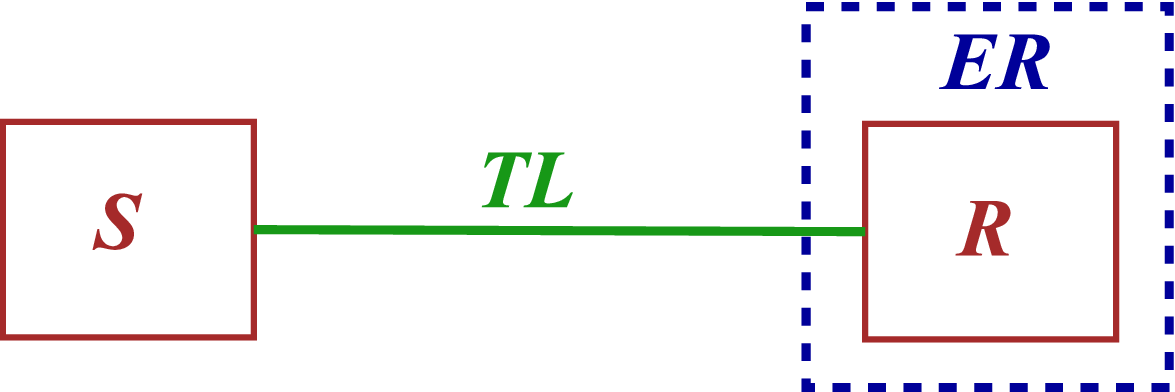}
\caption{The communication line including the sender $S$, receiver $R$ (embedded into the extended receiver $ER$) and transmission line. }
\label{Fig:line}
\end{figure}
The spin dynamics is governed by the XX-Hamiltonian with the dipole-dipole interaction:
\begin{eqnarray}\label{H}
H&=& \sum_{j>i}{D_{ij}(I_{ix}I_{jx}+I_{iy}I_{jy})},\\\label{com}
&& [H,I_{z}]=0, \;\;\;I_{z}=\sum_{i}{I_{iz}},
\end{eqnarray}
where $D_{ij}= {\gamma^2 \hbar}/{ r_{ij}^3}$ is the coupling constant between the $i$th and $j$th nodes,$ \gamma$ is the gyromagnetic ratio, $ r_{ij}$ is the distance between the $ith$ and $jth$ nodes, $ I_{i\alpha}(\alpha=x, y, z)$ is the projection operator of the $i$th spin on the $\alpha$-axis.
Due to the commutation condition (\ref{com}), the XX-Hamiltonian conserves the excitation number in a system and therefore has  the block diagonal form (\ref{Hd}) which, in our case of (0,1)-excitation spin dynamics,  reduces to two blocks:
\begin{eqnarray}
H={\mbox{diag}}(0,H_{1}).
\end{eqnarray}
The appropriate evolution operator $V(t)=e^{-i H t}$ has also the block-diagonal form,
\begin{eqnarray}
V(t)={\mbox{diag}}(1,V^{(1)}),\;\;\; V^{(1)}(t)=e^{-i \frac{H^{(1)}}{D_{12}} \tau}, \;\;\tau = D_{12} t,
\end{eqnarray}
where $\tau$ is the dimensionless time. 
At some time instant $\tau$, we apply a unitary transformation $U(\varphi)$ (which involves the set of free parameters $\varphi = \{\varphi_1,\varphi_2,\dots\}$) to the extended receiver. This transformation must conserve the excitation number, similarly to $V(t)$, i.e.,
\begin{eqnarray}
[I_{z}^{ER},U(\phi)]=0\;\;\; \Rightarrow \;\; U={\mbox{diag}}(1,U^{(1)}),
\end{eqnarray}
where $I _{z}^{ER}$ is the $z$-projection of the total spin moment in the state-space of the extended receiver $ER$.  Thus, the total transformation $W$ of the initial state is also a block-diagonal unitary transformation:
\begin{eqnarray}\label{W}
W(\tau)&=&\Big(I_{R,\overline{TL}}\otimes U\Big) V(\tau)={\mbox{diag}}(1,W^{(1)}(\tau)),\\\nonumber
&&W^{(1)}=\left(
\begin{array}{ccc}
W^{(1)}_{11}&\cdots&W^{(1)}_{1N}\cr
\cdots&\cdots&\cdots\cr
W^{(1)}_{N1}&\cdots&W^{(1)}_{NN}
\end{array}
\right)
\end{eqnarray}
where $I_{R,\overline{TL}}$ is the identity matrix in the space $R\cup \overline{TL}$,  $\overline{TL}$ means the $TL$ without nodes of the $ER$.

We represent $U^{(1)}$ in the exponential form,
\begin{eqnarray}
U^{(1)}(\varphi)=\prod_{j=1}^{N^{(ER)} (N^{(ER)}-1)} e^{iA_j(\varphi_j)},\;\; A_j^\dagger =A_j,
\end{eqnarray}
where each $A_j$ is the $N^{(ER)}\times N^{(ER)}$ Hermitian off-diagonal matrix which  involves only two nonzero elements introducing  the real parameter
$\varphi_j$. There are $N^{(ER)} (N^{(ER)}-1)$ linearly independent matrices of this type and the same number of the real parameters $\varphi$, 
\begin{eqnarray}\label{Udim}
\varphi=\{\varphi_i:\;i=1,\dots, N^{(ER)} (N^{(ER)}-1)\},
\end{eqnarray}
in the unitary transformation $U^{(1)}$.
Details of such parametrization  are given  in 
\cite{Z_2018}.

We consider the following tensor-product initial state
\begin{eqnarray}\label{ist}
s=\rho^{(S)}(0)\otimes \rho^{(TL,R)}(0),
\end{eqnarray}
where $\rho^{(S)}(0)$ is an arbitrary $(0,1)$-excitation space of $N^{(S)}$-qubit sender and $ \rho^{(TL,R)}(0)$ is the initial state of the joined subsystem $TL\cup R$ which is a $0$-excitation state:
\begin{eqnarray}
 \rho^{(TL,R)}(0) =|0\rangle_{R,TL} \; _{R,TL}\langle 0|.
\end{eqnarray}
The density matrix  of the receiver at some time instant $\tau$  is obtained from $\rho$ by partial tracing over $S$ and $TL$:
\begin{eqnarray}\label{rhot}
r(\tau)=Tr_{S,TL} \big(\rho(\tau)\big) = Tr_{S,TL} \big(W(\tau)\rho(0) W^{\dagger}(\tau)\big).
\end{eqnarray}
Both initial sender's state and the receiver's state have similar structures
 (compare with (\ref{rho}) and (\ref{rhoexpl})):
\begin{eqnarray}\label{s}
 s&=& \left(
\begin{array}{c|ccc}
s^{(0)}_{00}& s^{(1)}_{01}&\cdots& s^{(1)}_{0N^{(S)}} \cr\hline
s^{(-1)}_{10}&s^{(0)}_{11} &\cdots&s^{0}_{1N^{(S)}}\cr
\cdots & \cdots & \cdots & \cdots \cr
s^{(-1)}_{N^{(S)}0}&s^{(0)}_{N^{(S)}1} &\cdots&s^{0}_{N^{(S)}N^{(S)}}\end{array}
\right),\;\; \\\label{r}
r(\tau)&=& \left(
\begin{array}{c|ccc}
r^{(0)}_{00}(\tau)& r^{(1)}_{01}(\tau)&\cdots& r^{(1)}_{0N^{(R)}}(\tau) \cr\hline
r^{(-1)}_{10}(\tau)&r^{(0)}_{11}(\tau) &\cdots&r^{0}_{1N^{(R)}}(\tau)\cr
\cdots & \cdots & \cdots & \cdots \cr
r^{(-1)}_{N^{(R)}0}(\tau)&r^{(0)}_{N^{(R)}1}(\tau) &\cdots&r^{0}_{N^{(R)}N^{(R)}}(\tau)\end{array}
\right),\\\nonumber
&&
r^{(0)}_{ij}=\rho^{(R)}_{N-N^{(R)}+i,N-N^{(R)}+j},\;\;i,j=1,\dots,N^{(R)},\;\;
r^{(0)}_{00}=\rho^{(R)}_{00},\\\nonumber
&&
r^{(1)}_{0j}=\rho^{(R)}_{0,N-N^{(R)}+j},\;\;r^{(-1)}_{j0}=\rho^{(R)}_{N-N^{(R)}+j,0},\;\; j=1,\dots,N^{(R)},
\end{eqnarray}
and we set $N^{(S)}=N^{(R)}$ hereafter. We see that both $s$ and $r$ include $0$- and $\pm1$-order coherence matrices.

\section{Quantum state restoring}
\label{Section:StateRestoring}
Now we study 
the problem of state restoring \cite{FPZ_2021} in the case of $(0,1)$-excitation evolution in more details. 
We rewrite the definition (\ref{restore}) of the restored state for the $n$-order coherence matrices, $n=0,\pm 1$: 
\begin{eqnarray}\label{restored}
r^{(n)}_{ij}(t_0) = \lambda^{(n)}_{ij} s^{(n)}_{ij}, \;\;n=-1,0,1.
\end{eqnarray}
where the parameters $\lambda^{(n)}_{ij}$ are defined by the Hamiltonian and time instant $t_0$, but do not depend on the elements of $s$ and therefore they are universal for a given model of the communication line \cite{FPZ_2021}.  

It was shown in the case of complete  state space \cite{Z_2018} that there is a problem in restoring the diagonal elements of the density matrix. However, in the case of $(0,1)$-excitation spin dynamics this problem reduces to only one element $r^{(0)}_{00}$, which can not be restored. 
Moreover, we show that restoring the elements of the 1-order coherence matrix $r^{(1)}_{ij}$ automatically restores the elements $s^{(0)}_{ij}$, $i,j\neq0$. Of course, the elements of  $r^{(-1)}$ becomes also restored since $r^{(-1)} =\Big(r^{(1)} \Big)^\dagger$. 

We start with  writing the explicit relations between the elements of $r$ and $s$. For that,
we label the elements of $W^{(1)}$ by a triple index $\{n,m,l\}$, where the first, second and third  indexes are  withing the intervals, respectively,  $0\le n\le N^{(S)}$,  $0\le m\le N^{(TL)}$, $0\le l\le N^{(R)}$. Here $N^{(S)}$, $N^{(TL)}$ and $N^{(R)}$ are, respectively, the numbers of nodes in the sender, transmission line and receiver. Then Eq.(\ref{rhot}) yields 
\begin{eqnarray}\label{s00}
r^{(0)}_{00}&=&s^{(0)}_{00}+\sum_{{n,m}\atop{|n|+|m|=1}} \sum_{j,k=1}^{N^{(S)}} W^{(1)}_{n,m,0;j,0,0}s^{0}_{jk} (W^{(1)})^{\dagger}_{k,0,0; n,m,0} ,\\\label{s1}
r^{(1)}_{0n}&=&\sum_{j=1}^{N^{(S)}}  s^{(1)}_{0j}  (W^{(1)})^{\dagger}_{j,0,0;0,0,n}, \\\label{s0}
r^{(0)}_{nm}&=&\sum_{j,k=1}^{N^{(S)}} W^{(1)}_{0,0,n;j,0,0}s^{0}_{jk}  (W^{(1)})^{\dagger}_{k,0,0;0,0,m},\;\;n,m\neq 0. 
\end{eqnarray}
We see that restoring $r^{(1)}$ requires the ''diagonal'' form for $ W^{(1)}_{0,0,n;j,0,0}$:
\begin{eqnarray}\label{W1}
W^{(1)}_{0,0,n;j,0,0}= \lambda^{(1)}_{n0} \delta_{nj}.
\end{eqnarray}
This means that we have to find such parameters $\varphi$ of the unitary transformation $U^{(1)}$ that solve the following system of  $N^{(S)}(N^{(S)}-1)$
complex  equations
\begin{eqnarray}\label{Usolve}
W^{(1)}_{0,0,n;j,0,0} = 0,\;\; n\neq j,
\end{eqnarray}
 which is equivalent to the system of
\begin{eqnarray}\label{dW}
 2 N^{(S)}(N^{(S)}-1)
\end{eqnarray}
real equations.
Comparison of (\ref{Udim}) and (\ref{dW})  yields the condition on the lower bound for $N^{(ER)}$:
\begin{eqnarray}\label{Ub}
 N^{(ER)}(N^{(ER)}-1)>   2 N^{(S)}(N^{(S)}-1).
\end{eqnarray}

By virtue of (\ref{Usolve}), system (\ref{s00}) - (\ref{s0}) takes the following form
\begin{eqnarray}\label{s002}
r^{(0)}_{00}&=&s^{(0)}_{00}+ \sum_{{m=1}\atop{mn=0}}^{N^{(TL)}} \sum_{n,j=1}^{N^{(S)}} |W^{(1)}_{n,m,0;j,0,0}|^2 s^{0}_{jj}  ,\\\label{s12}
r^{(1)}_{0n}&=&(\lambda^{(1)}_{n0})^*  s^{(1)}_{0n} , \\\label{s02}
r^{(0)}_{nm}&=&\lambda^{(1)}_{n0}(\lambda^{(1)}_{m0})^* s^{0}_{nm},\;\; n,m>0,\;\; \Rightarrow\;\;\lambda^{(0)}_{nm}= \lambda^{(1)}_{n0}(\lambda^{(1)}_{m0})^*,
\end{eqnarray}
where $*$ means the complex conjugate.
Deriving (\ref{s002}) we use the unitarity of $W^{(1)}$.

\section{Entanglement in (0,1)-excitation state restoring}
\label{Section:Entanglement}
We consider two measures of entanglement. 

The first one is the measure of 
bi-qubit entanglement. We calculate the entanglement between any two qubits of  the  sender and its evolution to the entanglement between corresponding two qubits of the receiver. Using the Wootters criterion \cite{HW,Wootters} with the concurrence as a measure of quantum entanglement we show that the evolution of the concurrence reduces to its scaling with the scale factor independent on the initial sender's state. 

The second measure of entanglement is the double negativity (PPT-criterion)   \cite{Peres,VW}. It measures  the bi-partite entanglement between two multiqubit (in general) systems which are sender and receiver in our case.

\subsection{Scaling of pairwise entanglement}
We calculate the quantum entanglement by the Wootters criterion using the following formula \cite{HW,Wootters}:
\begin{eqnarray}\label{C}
C(r)&=&\max(0;\lambda_{1}-\lambda_{2}-\lambda_{3}-\lambda_{4}),\\\label{R}
R&=&\sqrt{r(\sigma_{y}\otimes\sigma_{y})r^{*}(\sigma_{y}\otimes\sigma_{y})},\;\;\sigma_{y}=\left(
\begin{array}{cc}
0&-i\cr
i&0
\end{array}
\right).
\end{eqnarray}
Here $\lambda_{i}, i=1,...,4$ are eigenvalues of the matrix $R$ and  $\lambda_{1}$ is the biggest of them.  

Consider the $(0,1)$-excitation density matrix  $s$ (\ref{s}). To calculate the concurrence between the $i$th and $j$th spins we have to calculate the partial trace of  $s$ with respect to all spins except the $i$th and $j$th to obtain the two-qubit density matrix $s^{(ij)}$, which we have to represent in the two-qubit basis 
\begin{eqnarray}
|00\rangle,\;\;|01\rangle,\;\;|10\rangle,\;\;|11\rangle.
\end{eqnarray}
 This matrix reads:
\begin{eqnarray}\label{block3}
&&s^{(ij)}=\left(
\begin{array}{cccc}
s_{00}&s_{0i}&s_{0j}&0\cr
s_{0i}^{*}&s_{ii}&s_{ij}&0\cr
s_{0j}^{*}&s_{ij}^{*}&s_{jj}&0\cr
0&0&0&0
\end{array}
\right),\;\;i,j>0.
\end{eqnarray}
It is remarkable that the nonzero eigenvalues of the matrix $R$ (\ref{R})  read
\begin{eqnarray}
\label{ev}
\lambda_{1}=\sqrt{s_{ii} s_{jj}} + |s_{ij}|,\;\; \lambda_{2}=\sqrt{s_{ii} s_{jj}} - |s_{ij}|,
\end{eqnarray}
where we take into account the non-negativity of the density matrix $s$, so that
\begin{eqnarray}
s_{ii}s_{jj}-|s_{ij}|^2\geq 0.
\end{eqnarray}
Then,  concurrence between the $i$th and $j$th nodes of the sender reads 
\begin{eqnarray}\label{C2}
C_{ij}^{(s)}=2|s_{ij}|,\;\; i\neq j.
\end{eqnarray}
We shall emphasize that the concurrence does not depend on the  elements of the $\pm1$-order coherence matrices of $s^{(ij)}$. Thus we proved the following Proposition.

{\bf Proposition 1.} The concurrence of 2-qubit state  (\ref{block3}) is defined by the single element of the density matrix $s$ by formula (\ref{C2}).

{\bf Proposition 2.} If all non-diagonal elements in the 0-order coherence  matrix $r^{(0)}$ of  state $r$ (\ref{r})  are proportional to the appropriate elements of $s^{(0)}$ of state $s$ (\ref{s}) according to (\ref{restored}),
then 
\begin{eqnarray}\label{rs}
\frac{C^{(r)}_{ij}}{C^{(s)}_{ij}}=2|\lambda^{(0)}_{ij}|= 2 | \lambda^{(1)}_{i}(\lambda^{(1)}_{j})^*|,\;\; \forall \;i,j, \;\;i\neq j.
\end{eqnarray}

It is important that relation (\ref{rs}) holds for all pairs $(i,j)$ of different spins of the sender (and receiver)  in the case of state restoring. 
Therefore, the long-distance state restoring scales the pairwise concurrence with the universal (i.e., independent on the sender's initial state $s$) scale parameter $2|\lambda^{(0)}_{ij}|$. This scale parameter is defined by the Hamiltonian which governs the spin dynamics, and by the  time instant $\tau$ selected  for state registration. 


\subsection{$S-R$ entanglement via PPT criterion}
The Wootters criterion is well applicable  for calculating the 2-qubit entanglement. However, to calculate the entanglement between two multi-qubit subsystems another method must be implemented. We consider the PPT criterion \cite{Peres,VW} for this purpose.
According to this criterion, the entanglement between two quantum subsystems $S$ and $R$ of system is expressed in terms of the so-called double negativity $N_{SR}$, which is the doubled  sum of the absolute values of the negative eigenvalues 
$\lambda_{i}<0$ of the matrix $(\rho^{(R S)})^{T_{S}}$,
\begin{eqnarray}
N_{SR}= 2\sum_{\lambda_{i}<0}|\lambda_{i}|.
\end{eqnarray} 
Here
\begin{eqnarray}
\rho^{(R S)}(\tau) = {\mbox{Tr}}_{TL}\; \rho(\tau),
\end{eqnarray}
and the superscript $T_{S}$ means the partial transpose with respect to the subsystem $S$.

\section{Examples: structural restoring of 2- and 3-qubit states. }
\label{Section:Examples}
\subsection{Time instant for state registration.}
First, we fix the time instant $\tau$   that maximizes the minimal of the scale parameters $\lambda^{(1)}_{i0}$, $i=1,\dots,N^{(S)}$, which we denote $\lambda(N)$. At that, we fix the optimal   parameters $\varphi$ obtained as solutions  of Eq.(\ref{Usolve}). In other words, we define the parameter $\lambda(N)$ according to the following scheme. 

\begin{enumerate}\item
For each chain length $N$, we scan the time interval $0\le \tau\le T\gtrsim N$.
\item
At each $\tau$-instant, we find 1000 independent solutions $\varphi^{(i)}(\tau)$ of (\ref{Usolve}), $i=1,\dots,1000$.
\item
For each solution $\varphi^{(i)}(\tau)$,  we  calculate $\lambda^{(1)}_{j0}(\tau,\varphi^{(i)})\equiv \lambda^{(1;i)}_{j0}$, $j=1,\dots,N^{(S)}$.
\item
Find $\lambda_{min}^{(i)}(\tau)=\min\Big(\lambda^{(1;i)}_{1}(\tau),\dots, \lambda^{(1;i)}_{N^{(S)}}(\tau)\Big)$.
\item
Then 
\begin{eqnarray}\label{lamN}
\lambda(N)= \max_{\tau,i} \;\lambda_{min}^{(i)}(\tau). 
\end{eqnarray}
 We denote the  time instant  yielding maximum in (\ref{lamN}) by  $\tau(N)$.
\end{enumerate}

\subsection{Two-qubit state restoring.}
We consider the two-qubit receiver ($N^{(S)}=N^{(R)}=2$) with the three-qubit extended receiver ($N^{(ER)}=3$). In the above protocol of calculating $\lambda(N)$, we include the set of 1000 solutions 
$\varphi^{(i}$ of system (\ref{Usolve}), ${i=1,\dots,1000}$.
The graphs  $\lambda(N)$ and  $\tau(N)$ are  shown in Fig.\ref{Fig:mpr}. 
Notice  that   $\tau(N)$ appears to be a straight line.}

\begin{figure}[ht]
\centering
  \begin{subfigure}[c]{0.45\textwidth}
    \centering
    \includegraphics[width=\textwidth]{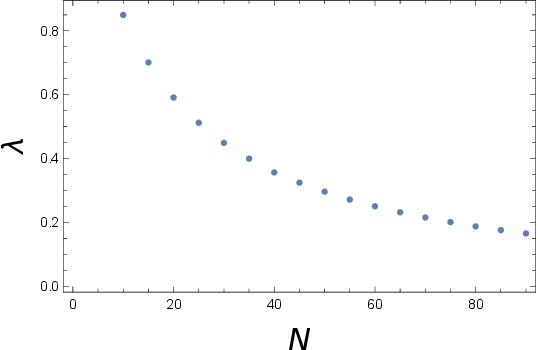}
    \caption{}
  \end{subfigure}
\hfill
\centering
  \begin{subfigure}[c]{0.45\textwidth}
    \centering
    \includegraphics[width=\textwidth]{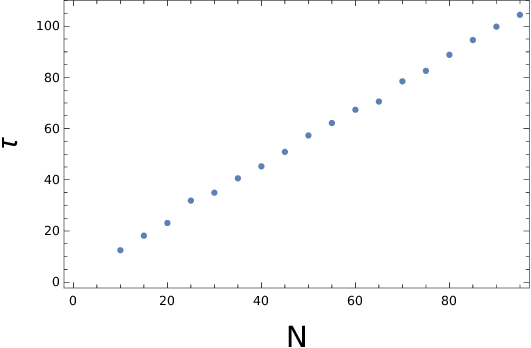}
    \caption{}
  \end{subfigure}
\caption{The parameters $\lambda(N)$   (a) and $\tau(N)$ (b) as  functions of the chain length $N$ for $N^{(S)}=2$, $N^{(EX)}=3$. }
\label{Fig:mpr}
\end{figure}

Let us consider a particular example of two-qubit state restoring in the communication line of $N=10$ nodes. 
An optimized unitary  transformation  at ${\tau=12.5}$ reads 
\begin{eqnarray}\label{block}
U^{(1)}=\left(
\begin{array}{cccc}
0.008 e^{i2.790}&0.967 e^{-i 3.109}&0.052 e^{i 1.132}&0.249 e^{i 3.058}\cr
0.819 e^{i 2.649}&0.027 e^{-i 1.903}&0.530 e^{-i 2.974}&0.215 e^{i 1.511}\cr
0.195 e^{-i 2.276}&0.249 e^{i 2.015}&0.264 e^{i 2.015}&0.911 e^{-i 2.031}\cr
0.538 e^{-i 1.225}&0.039 e^{-i 1.332}&0.804 e^{i  1.470}&0.249 e^{i1.001}
\end{array}
\right).
\end{eqnarray}
Corresponding matrix $r$ reads
\begin{eqnarray}\label{block2}
&&r=\left(
\begin{array}{ccc}
r_{00}&0.655 e^{-i 0.094}s_{01}&0.584 e^{i1.783}s_{02}\cr
0.655 e^{i 0.094}s_{01}^*&0.429s_{11}&0.382 e^{i1.876}s_{12}\cr
0.584 e^{-i 1.783}s_{02}^*&0.382 e^{-i 7.876 }s_{12}^*&0.341s_{22}\cr
\end{array}
\right),\\\nonumber
&&\lambda^{(1)}_1=0.655 e^{-i 0.094},\;\;\lambda^{(1)}_2= 0.5834 e^{i1.783}.
\end{eqnarray}
where $r_{00}=s_{00} + 0.571 s_{11} + 0.659 s_{22}$. 

\begin{figure}[ht]
\includegraphics[scale=1]{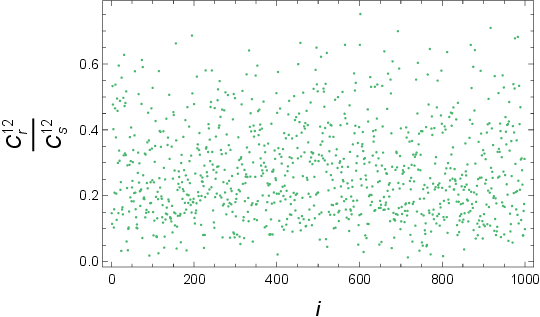}
\caption{The ratio $C^{(r)}_{12}/C^{(s)}_{12}$ in dependence on  number of solution  $\varphi^{(i)}$, $i$; the maximal value of this ratio
equals 0.752.}
\label{Fig:mpr2}
\end{figure}

 Now we turn to the bi-particle entanglement.
To illustrate the dependence of  the ratio 
$C^{(r)}_{12}/C^{(s)}_{12}$ on the solution $\varphi^{(i)}$ of  system (\ref{Usolve}), we plot  $C^{(r)}_{12}/C^{(s)}_{12}$  for different  solutions $\varphi^{(i)}$, $i=1,\dots,1000$  and chain length ${N=10}$ at ${\tau=12.5}$, see Fig.\ref{Fig:mpr2}. The maximal value of this ratio is 0.752. The maximal density of the points on Fig.\ref{Fig:mpr2} is observed at $C^{(r)}_{12}/C^{(s)}_{12} \sim 0.1 \div 0.3$.

\subsection{3-qubit state restoring.}
As another example, we investigate and optimize (over 1000 solutions $\varphi^{(i)}$ of system (\ref{Usolve})) of  the ratios $C^{(r)}_{ij}/ C^{(s)}_{ij} = 2|\lambda^{(0)}_{ij}|$  for three possible
pairs of spins in a three-qubit receiver and sender.  The length of the chain is $N=10$, the extended receiver includes $5$ qubits. The time corresponding to the maximal  ratio (or maximal $|\lambda^{(0)}_{ij}|$) is found for each pair of qubits. The results are collected in Table \ref{Table:1}.
\begin{table}
\begin{tabular}{ |c | c | c| }
\hline
 $\{i,j\}$ & $\tau$& $C^{(r)}_{ij}/ C^{(s)}_{ij}$ \\ 
\hline
 \{1,2\} & 110&  0.092 \\
\hline  
\{2,3\} &108.5 & 0.054\\
\hline     
\{1,3\}& 109.4& 0.025\\
\hline
\end{tabular}
\caption{The ratio $C^{(r)}_{ij}/ C^{(s)}_{ij}$ maximized over 1000 $\varphi^{(i)}$ and over time. 
The optimal  time instant $\tau$ for each pair $\{i,j\}$ is given.}
\label{Table:1}
\end{table}

\subsubsection{$S-R$ entanglement  via PPT criterion.}
\begin{figure}[ht]
\includegraphics[ scale=0.7]{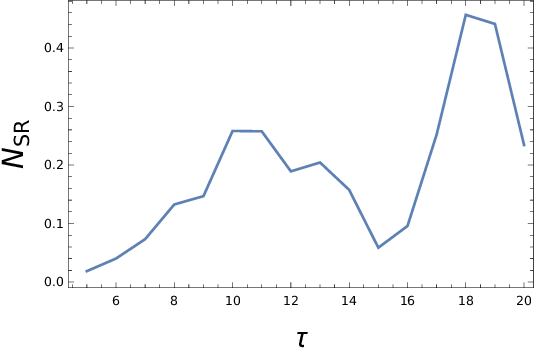}
\caption{The $\tau$-dependence of the double negativity averaged over 200 arbitrary pure initial states (\ref{Psi}).}
\label{Fig:mpr3}
\end{figure}

Now we strive to maximize entanglement between $S$ and $R$ for this example using the PPT criterion (double negativity $N_{SR}$ is a measure of entanglement).
First, for each $\tau$, $0\le\tau\le 2 N=20$, we optimize  $\varphi$  calculating $\lambda(N)$, $N=10$, (see Eq.(\ref{lamN})) with maximization over 500 solutions $\varphi^{(i)}$, $i=1,\dots,500$ of system (\ref{Usolve}). Then, for the optimized $\varphi$, we calculate double negativity $N_{SR}$  averaged over $200$  pure sender's initial states taken in the form
\begin{eqnarray}\label{Psi}
|\Psi^{(S)}\rangle &=& \sum_{i=0}^3 a_i |i\rangle,\\\nonumber
&& a_0 = \cos\,\psi_1 \;\cos\,\psi_2\;\cos\,\psi_3,\\\nonumber
&&
a_1 = \sin\,\psi_1 \;\cos\,\psi_2\;\cos\,\psi_3 \; e^{i \phi_1}\\\nonumber
&&
a_2 = \sin\,\psi_2 \;\cos\,\psi_3\;e^{i \phi_2} \\\nonumber
&&
a_3 = \sin\,\psi_3 \;e^{i \phi_3},
\end{eqnarray}
where $\psi_i$ and $\phi_i$ are randomly chosen real parameters. 
The  graph of $N_{SR}$ as a function of $\tau$ calculated in this way is shown  on Fig.\ref{Fig:mpr3}. Two maxima at $\tau\sim 10$ and $\tau\sim 18$ are observed.

\begin{figure}[ht]
\includegraphics[ scale=0.7]{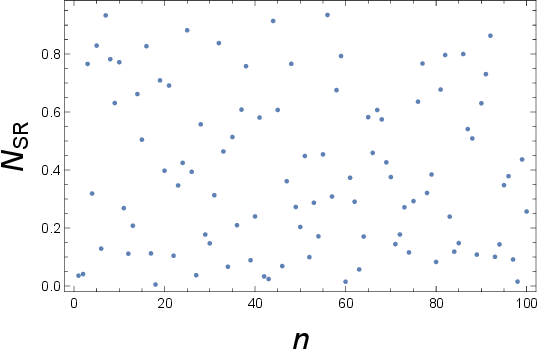}
\caption{The diagram of $S-R$ entanglement in dependence on different sender's initial states numbered by $n$ at $\tau=18$; the maximal value of this ratio  equals $N_{SR}=0.935$.}
\label{Fig:mpr4}
\end{figure}

The distribution of double negativity $N_{SR}$ for 100 random sender's initial states at 
$\tau=18$ (the second maximum on Fig.\ref{Fig:mpr3}) is shown in Fig.\ref{Fig:mpr4} with the  maximal value is $N_{SR}=0.935$

\section{Conclusions}
\label{Section:Conclusions}
We consider the process of remote structural state restoring of (0,1)-excitation states (i.e., 
states including only 0- and 1-excitations). We show, that in this case almost complete  structural  state restoring can be performed by restoring only the 1-order coherence matrix. At that, the only non-restorable element of the density matrix is the element from the 0-excitation block of 0-order coherence matrix. 

We study the entanglement in this process and find out that the concurrence between any two sender's qubits scales to the
concurrence between the corresponding  two qubits of the receiver. It is remarkable that the factor of such scaling is universal, i.e., it does not depend on the particular sender's state. This factor is defined by the Hamiltonian and time instant for state registration. Examples of 2 and 3-qubit state restoring are presented.

Two above results (almost complete structural state restore and scaled concurrence)  can be referred to the features of (0,1)-excitation evolution which can not be observed in the case of evolution in  the complete state space.

The entanglement between the sender and receiver is studied using the PPT-criterion (doble negativity). In this case we consider the $N=10$ chain with 3-qubit receiver (sender) and 
average the double negativity over 200 random initial pure states of the sender. We  define the maxima at $\tau\sim 10$ ($\sim N$) and $\tau\sim 18$ ($\sim 2 N$).  

{\bf Acknowledgments.} The work was performed as a part of a state task, State Registration
No. AAAA-A19-119071190017-7.

\end{document}